\documentclass[a4paper,showpacs,twocolumn,amsmath,amssymb,superscriptaddress,floatfix,prb,preprintnumbers,footinbib]{revtex4-1}

\usepackage[utf8]{inputenc}
\usepackage{graphicx}
\usepackage{subfigure}
\usepackage[colorlinks=true,citecolor=blue]{hyperref}
\usepackage{bm}

\begin{document}

\title{Mesoscopic Diffusion Thermopower in Two-Dimensional Electron Gases}

\author{Stephan Rojek} 
\author{J\"urgen K\"onig} 
\affiliation{Theoretische Physik and CENIDE, Universit\"at Duisburg-Essen, 47048 Duisburg, Germany}

\date{\today}

\begin{abstract}
The diffusion of energy that is locally deposited into two-dimensional electron gases by Joule heating generates transverse voltages across devices with broken symmetry.
For mesoscopic structures characterized by device dimensions comparable to the energy diffusion length, the resulting thermopower strongly depends on details of the potential profile defined by electric gates.
We discuss these mesoscopic features within a diffusion thermopower model and propose schemes to measure the energy diffusion length and its dependence on gate voltage. 
\end{abstract}


\maketitle
\section{Introduction}
The principle of thermoelectric devices is the conversion of heat to charge currents and vice-versa.\cite{tritt2011,white_energy-harvesting_2008}
For practical applications, a key question is how to achieve this conversion in the most efficient way.
In addition, there is fundamental interest in studying thermoelectric effects since they provide insight into underlying transport mechanisms. 
The information derived from heat and charge transport is beyond that contained in measurements of electric conductivity.
This especially applies to mesoscopic devices,\cite{giazotto2006,Kantser2006} where sample size and geometry affect transport, while, for bulk systems, only a few material parameters are relevant.   
Two-dimensional electron gases (2DEGs) are particularly suited for fundamental studies because the variety of possible structures and the possibility to tune the carrier density by gate voltages allows for controlling mesoscopic aspects of charge and heat transport.
In particular, one can reach the regime in which an important length scale, the energy diffusion length, becomes comparable to the system size, such that mesoscopic effects for heat transport can be expected. 

For 2DEGs realized in a semiconductor heterostructure, the momentum-relaxation time, $\tau_\text{e}$, that determines the mobility of the charge carriers, is dominated by elastic scattering processes for typical low-temperature measurements. 
The corresponding elastic scattering length or elastic mean-free path $l_{\rm e}=v_{\rm F} \tau_{\rm e}$, where $v_{\rm F}$ is the Fermi velocity, marks the separation between ballistic and diffusive transport of the electrons.
For device dimensions $L$ larger than $l_{\rm e}$, transport is diffusive with diffusion constant $D=v_{\rm F}^2 \tau_{\rm e}/2$ (in two dimensions).
An important length scale for heat transport results from the inelastic or energy-relaxation time, $\tau_{\rm i}$, given by scattering processes in which the energy transfer between electrons and lattice exceeds $k_\text{B}T$.
This introduces the energy diffusion length $l=\sqrt{D\tau_{\rm i}}$ as a length scale, on which the local energy density of the charge carrier system varies spatially in a stationary situation.\cite{bass,anderson,lohvinov,ivchenko,Olbrich2011}

Since at low temperature $\tau_{\rm e} \ll \tau_{\rm i}$, it is possible to realize devices with $l_{\rm e} \ll L \lesssim l$, for which heat transport behaves mesoscopically while charge transport does not.
This regime, that we want to address in this paper, is distinctively different from macroscopic thermoelectric devices but also from nanoscale systems,\cite{staring,dzurak,llaguno,scheibner1,svensson,entin1,juergens,bakker,sothmann,mazza} in which energy quantization and Coulomb charging are important, and from the ballistic or quasiballistic regime,\cite{washburn,marcus,predel,fischer,javey,predel,buhmann,hua} in which charge transport behaves mesoscopically.\cite{wang}
Our calculations are based on a noninteracting-electron picture.
This is in contrast to recent experimental studies of diffusion thermopower in strongly correlated systems, e.g., those displaying the fractional quantum Hall effect\cite{Possanzini2003,Faniel2005,Chickering2010} or low-concentration samples with unconconventional metallic phases.\cite{Moldovan2000,Goswami2009,Narayan2012}

\begin{figure}
\includegraphics[width=.49\textwidth]{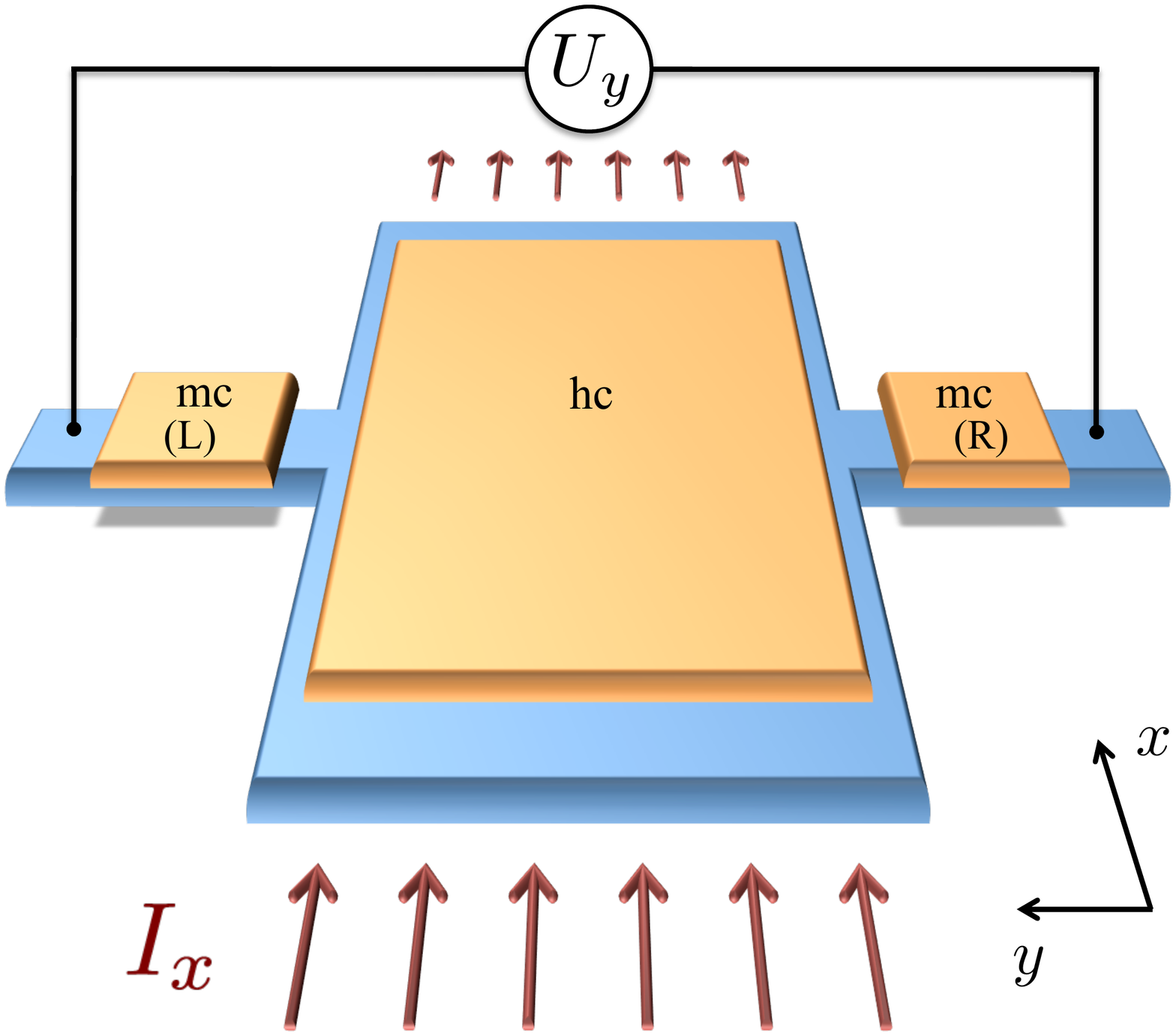}
\caption{\label{fig:setup}(Color online) 
Schematic picture of a transverse thermoelectric rectifier.
The carrier density of a 2DEG (blue region) is given by the potentials $V_\text{mc}^\text{L}$, $V_\text{mc}^\text{R}$, and $V_\text{hc}$ locally controlled by voltages applied to top gates.
An applied current $I_x$ heats up the electrons in the heating channel (hc).
The transverse, thermoelectric output voltage $U_y$ depends on the gate voltages applied to the gates on top of the modulation channels (mc) on the left (L) and right (R) hand side.}
\end{figure}

A temperature gradient can drive an electric current, or, for open electric contacts, generate an electric-voltage gradient.
The strength of this thermopower is characterized by the Seebeck coefficient $S=U/\Delta T$, where $U$ is the voltage generated by a temperature difference $\Delta T$ between two contacts.
The Seebeck coefficient is a useful quantity whenever the temperature of the two contacts are given as a boundary condition.
While this is the case in many experimental setups involving macroscopic systems, the situation may be different for mesoscopic samples in which temperature gradients appear as a consequence of Joule heating by local electric currents. 
To be more specific, we will consider devices as sketched in Fig.~\ref{fig:setup}: metallic gates, with a length in $x$-direction that is much larger than the width, modulate the electric potential of a 2DEG in $y$-direction. 
Thereby, an electric current $I_{x}$ driven through the 2DEG in $x$-direction heats up the electron system which, in turn, generates a perpendicular output voltage $U_y$.

In such a mesoscopic all-electric setup, heat generation and heat diffusion have to be treated on the same footing, i.e., the local temperature is not a priori known but needs to be determined self-consistently.
It is, therefore, more natural to characterize the thermopower by relating the output voltage $U_y$ to the input heating current or voltage rather than a temperature difference.
As we will see below, a convenient measure of the device's performance is the dimensionless quantity 
\begin{equation}
	s = \frac{eU_y}{\omega}
\end{equation}
where $\omega$ is the application of energy per electron provided by the heating current.
For devices with all relevant geometric lengths exceeding the energy diffusion length, $s$ is just a number determined by material parameters.
In the mesoscopic regime, however, $s$ will depend on various geometric lengths such as the width of regions heated by an applied current (heating channel), the width of regions with a modulated charge density (modulation channel), the distance between heating and modulation channels, and the width of the potential steps at the edges of modulation channels.

The aim of this paper is to discuss the influence of the energy diffusion length on the thermoelectric output voltage generated by input electric heating currents for devices as sketched in Fig.~\ref{fig:setup} and experimentally realized in, e.g., Refs.~\onlinecite{gallagher1990,Chickering2009,ganczarczyk_transverse_2012}.
For practical reasons, the voltage contacts are spatially separated from the heating channel(s), i.e., the voltage contacts remain at base temperature.
Therefore, a finite output voltage can only occur if the reflection symmetry about the $x$-axis is broken, e.g., by placing a modulation channel on one side only.
Since the unipolar output voltage, $U_y$, is independent of the input current's direction, the studied setups effectively act as transverse rectifiers.
The underlying mechanism of this rectification, diffusive thermopower, is different from ballistic rectification in samples with dimensions small compared to the elastic mean free path $l_{\rm e}$ which include symmetry-breaking scatterers\cite{Song1998,Matthews2012} or asymmetric cross junctions.\cite{Knop2006,salloch2010}  
It is, furthermore, different from hot-electron thermopower of quantum point contacts,\cite{Molenkamp1990a,Wiemann2010} which sometimes enhances the rectification signal in ballistic rectifiers.\cite{Salloch2009,salloch2011} Another source of a nonlinear output signal is related to the number of current-carrying modes in channels connecting input and output terminals. An asymmetry in the mode numbers may be influenced by the position of a central scatterer and/or gate voltages. This and related mechanisms have been discussed for both four-\cite{fleischmann,haan1,haan2,Hackens2004} and  three-terminal\cite{Hieke2000,xu_electrical_2001,Shorubalko2001,Worschech2001,irie,jordan2} devices in the range from diffusive to quantized transport.

\section{Diffusion Thermopower Model\label{Theo}}

The diffusion thermopower model is based on a semiclassical approach to thermoelectric transport.\cite{Ashcroft1976}
An applied electric current provides additional energy for the electronic system  within the heating channel (the case of more than one heating channel is addressed below). 
The excess energy diffuses into perpendicular direction (as indicated in Fig.~\ref{fig:setup}, we assume translational invariance in $x$-direction, i.e., all quantities only depend on the position in $y$-direction).
In steady state, the distribution of the excess energy is described by a temperature profile $T(y)$ with a maximum somewhere in the heating channel.
For spatially well-separated voltage contacts, the temperature decreases towards the base temperature of the lattice, $T(y) \rightarrow T_0$ for $y\rightarrow \pm \infty$.
The local temperature gradient induces a voltage gradient proportional to the Seebeck coefficient $S(y)$, which yields the total voltage drop
\begin{align}
	U_{y}= \int^\infty\limits_{-\infty} \text{d}y\ S(y)\  \partial_y T(y)\; . \label{MottInt}
\end{align}
The contributions from regions with positive and negative temperature gradients compensate each other for a left-right symmetric setup. 
In this case, $U_y$ vanishes.
An applied gate voltage on either side of the heating channel breaks this symmetry by modulating the charge density, leading to a finite $U_y$.
The density modulation changes the local Seebeck coefficient $S(y)$ that explicitly enters Eq.~(\ref{MottInt}).
Moreover, the heat diffusion both within the electronic system and from electrons to the lattice are modified by the gate voltage via the energy dependence of $\tau_\textrm{e}$ and $\tau_\textrm{i}$, which, in turn, affects the temperature profile and, thus, indirectly $U_y$.

For determining the temperature profile in the most general case, we extend the derivation presented in Ref.~\onlinecite{ganczarczyk_transverse_2012} by allowing for an arbitrary $y$-dependence of the potential profile.
This extension becomes relevant, e.g., when the changes in the potential profile at the edges of the modulation channels are not sharp but smeared on a scale comparable to the energy diffusion length.

\subsection{Heat-Balance Equation}

To derive the temperature profile, we identify the processes which change the local heat density in the electronic system.
First, heat is generated by Joule heating $\mathbf{j} \cdot \mathbf{E}$ (power per area), where $\mathbf{j}$ is the charge current per unit length in the 2DEG and $\mathbf{E}$ the electric field.
Second, heat is transferred from the electronic system to the lattice, which we treat as a reservoir with fixed temperature $T_0$.
Phenomenologically, this process can be modeled by $c_V (T-T_0) /\tau_\textrm{i}$ for small temperature difference $\delta T= T-T_0$, where $c_V$ is the heat capacity per unit area and $\tau_\textrm{i}$ the energy-relaxation time of the electrons taken at the Fermi energy level.\cite{anderson,hansch}
The latter accounts for scattering processes with energy transfer of $k_\text{B}T$ or more, which, at low temperature, are mainly given by scattering with acoustic phonons.~\cite{Kawamura1992} 
Scattering processes with smaller energy transfer do not affect the local temperature but, still, charge and energy diffusion and, thereby, the charge, $\mathbf{j}$, and heat current, $\mathbf{j}^q$, per unit length.
As a result, we get
\begin{align}
	\nabla \cdot {\mathbf{j}^q} = \mathbf{j} \cdot \mathbf{E}  -c_V \frac{\delta T}{\tau_\textrm{i}} \, ,
\label{continuity}
\end{align}
for the heat-balance equation.

Translation invariance in $x$-direction implies $\mathbf{j}^q=\mathbf{j}^q(y)$ and, thus, $\nabla \cdot {\mathbf{j}^q} = \partial j_y^q / \partial y$.
Furthermore, for open voltage contacts electric current is flowing in $x$-direction only, i.e., $\mathbf{j}=(j_x,0,0)$, which yields the Joule heating  $\mathbf{j} \cdot \mathbf{E} = j_x E_x$, where $E_x$ is the electric field that drives the electric current in $x$-direction.
To eliminate the current densities in Eq.~(\ref{continuity}), we make use of the linear-response relations
\begin{align}
 j_x &= \sigma\, E_x \label{eq:jx}\, ,\\
 j_y^q &= -\kappa\ \frac{\partial T}{\partial y}\, ,
\end{align} 
which define the (two-dimensional) electric and thermal conductivity $\sigma$ and $\kappa$, respectively.
That leads to a differential equation for the temperature profile
\begin{align}
	T- \frac{l^2}{\kappa} \frac{\partial}{\partial y} \left(\kappa \frac{\partial T}{\partial y} \right)= T_\textrm{bulk} \, ,
	\label{Tdiff}
\end{align}
where the energy diffusion length
\begin{align}
	l =\sqrt{\frac{\kappa \tau_\textrm{i}}{c_V}}
\end{align}
defines the length scale on which the temperature profile varies (see Refs.~\onlinecite{lohvinov,ganczarczyk_transverse_2012} for diffusive temperature differential equations in case of a constant carrier density). 
The inhomogeneity
\begin{align}
\label{tbulk}
	T_\textrm{bulk} = T_0 + \frac{\omega}{c_V/n}
\end{align}
is the temperature the electrons would acquire in a bulk sample much larger than the energy diffusion length $l$.
Here, $n$ is the electron density and
\begin{align}
\label{om_gen}
	\omega = \frac{\sigma \tau_\textrm{i} E_x^2}{n}
\end{align}
is readily identified as the energy increase per electron provided by the heating current ($\sigma E_x^2$ is the power density due to Joule heating and $\tau_\textrm{i}$ is the characteristic time on which the electrons gain energy before they are scattered inelastically).
We remark that $l(y)$, $\kappa(y)$, $\sigma(y)$, and $E_x(y)$ are, in general, $y$-dependent. 
In the case of $l$, $\kappa$, and $\sigma$ that results from the charge-density modulation. 
The externally applied electric field, on the other hand, is constant within the heating channels, but vanishes outside.
Any temperature dependence of $l$, $\kappa$, and $\sigma$ can be neglected in Eq.~\eqref{Tdiff} because we restrict ourselves to linear order in $\delta T$ and quadratic order in $E_x$.

\subsection{Boundary Conditions\label{bound}}

Since the differential equation for the temperature, Eq.~\eqref{Tdiff}, is of second order, two boundary conditions are needed.
Far away from the heating channels, $y \rightarrow \pm \infty$, the temperature remains at base temperature, $T\rightarrow T_0$.

Sharp edges of the modulation and the heating channels divide the integration range due to discontinuities in the potential or the inhomogeneity $T_\text{bulk}$. 
In this case, the solutions of the differential equation on the left and on the right of the interface need to be matched.
This is achieved by making use of the two conditions that the heat current entering the interface from one side is (i) equal to the heat current leaving the interface from the other side and (ii) equal to the heat current flowing through the interface.
The first condition leads to the continuity of  $\kappa (\partial T/\partial y)$.

The heat current through the interface, which enters the second condition, is driven by a temperature difference across the interface.
(This discontinuity of the temperature at the interface is accompanied by a jump in the electrochemical potential since for open voltage contacts there is no charge current flowing through the interface.)
As derived in Ref.~\onlinecite{ganczarczyk_transverse_2012}, the size of this temperature jump is, in general, much smaller than the temperature variation within the 2DEG and can, thus, be neglected.
This can be interpreted in the following way: 
The heat current through the interface is carried by ballistic electrons and, therefore, only limited by the finite transmission probability $|t|^2$ for an incoming electron to pass the interface.\cite{sivan_multichannel_1986,laikhtman,dugaev}
The heat conductivity in the 2DEG, on the other hand, is limited by elastic scattering, characterized by the scattering length $l_\textrm{e}$.
The length scale associated with the temperature gradient driving the heat current, however, is the energy diffusion length $l$ and, thus, much larger.
As a consequence, the temperature jump at the interface is parametrically given by a factor of $l_\textrm{e}/(l |t|^{2})$ smaller than the temperature change in the 2DEG and, therefore, negligible (unless $|t|^{2} \lesssim l_\textrm{e}/l$).
In conclusion, we demand, as the second matching condition, the temperature $T$ to be continuous at the interface.

\subsection{Sommerfeld Expansion\label{sommerfeld}}

The thermopower diffusion model presented above contains the linear transport coefficients $\sigma$ and $\kappa$ and the specific heat capacity $c_V$.
At low temperature, $k_\text{B}T_0 \ll \epsilon_\text{F}$, it is sufficient to determine them to lowest order in the Sommerfeld expansion.\cite{Ziman,Ashcroft1976}
We express the results by making use of the density of states $\rho=m^\ast/\pi\hbar^2$ for free electrons with effective mass $m^\ast$ and of the diffusion constant $D=\epsilon_{\rm F} \tau_{\rm e} (\epsilon_{\rm F})/m^\ast$ at the Fermi energy (both for two-dimensional systems), and obtain
\begin{subequations}
\label{Sommerfeld_1}
\begin{align}
	\sigma &=  e^2 \rho\, D \label{sigma}\, , \\
	\kappa &= \rho  \frac{\pi^2}{3} k_\text{B}^2T_0\, D\, , \label{kappa} \\
	c_V &=  \kappa/D  
	\label{c_V} \, ,
\end{align} 
\end{subequations}
and from this, we derive 
\begin{subequations}
\label{Sommerfeld_2}
\begin{align}
	l &= \sqrt{D\, \tau_i } = \left.\sqrt{\frac{\epsilon \tau_\text{e}(\epsilon)\tau_\text{i}(\epsilon) }{m^\ast}} \right |_{\epsilon=\epsilon_\textrm{F}}  \label{ediff}\, , \\
 	S &=-\frac{\pi^2}{3}\frac{k_\text{B}^2T_0}{e}\,\frac{\partial \ln \left( \epsilon \tau_\textrm{e}(\epsilon) \right)}{\partial\epsilon}  \bigg|_{\epsilon=\epsilon_\textrm{F}}  \label{thermopower} \, ,\\
 	\omega &=\frac{(eE_x l)^2}{\epsilon_{\rm F}} \, . 
\label{om_som}
\end{align} 
\end{subequations}
Equation~\eqref{thermopower} is the Mott formula for the thermopower in a two-dimensional electron gas.\cite{Cutler1969} 
The $y$-dependent charge carrier density enters $l$, $S$, and $\omega$ via the Fermi energy $\epsilon_\text{F}(y)$, measured relatively to the lower subband edge. 
In order to express all results in dimensionless quantities, we choose the case of vanishing gate voltages (which yields $V_\text{hc}=V_\text{mc}=0$) as a reference, and denote the corresponding Fermi energy by $\epsilon_0$.
Similarly, $l$, $n$, and $\omega$ taken at this energy fixes the reference quantities $l_0$, $n_0$, and $\omega_0$, and $s_0$ is defined as $eU_y/\omega_0$. 

\subsection{Gate-Voltage Dependent Thermopower}

\begin{figure}
	\includegraphics[width=.49\textwidth]{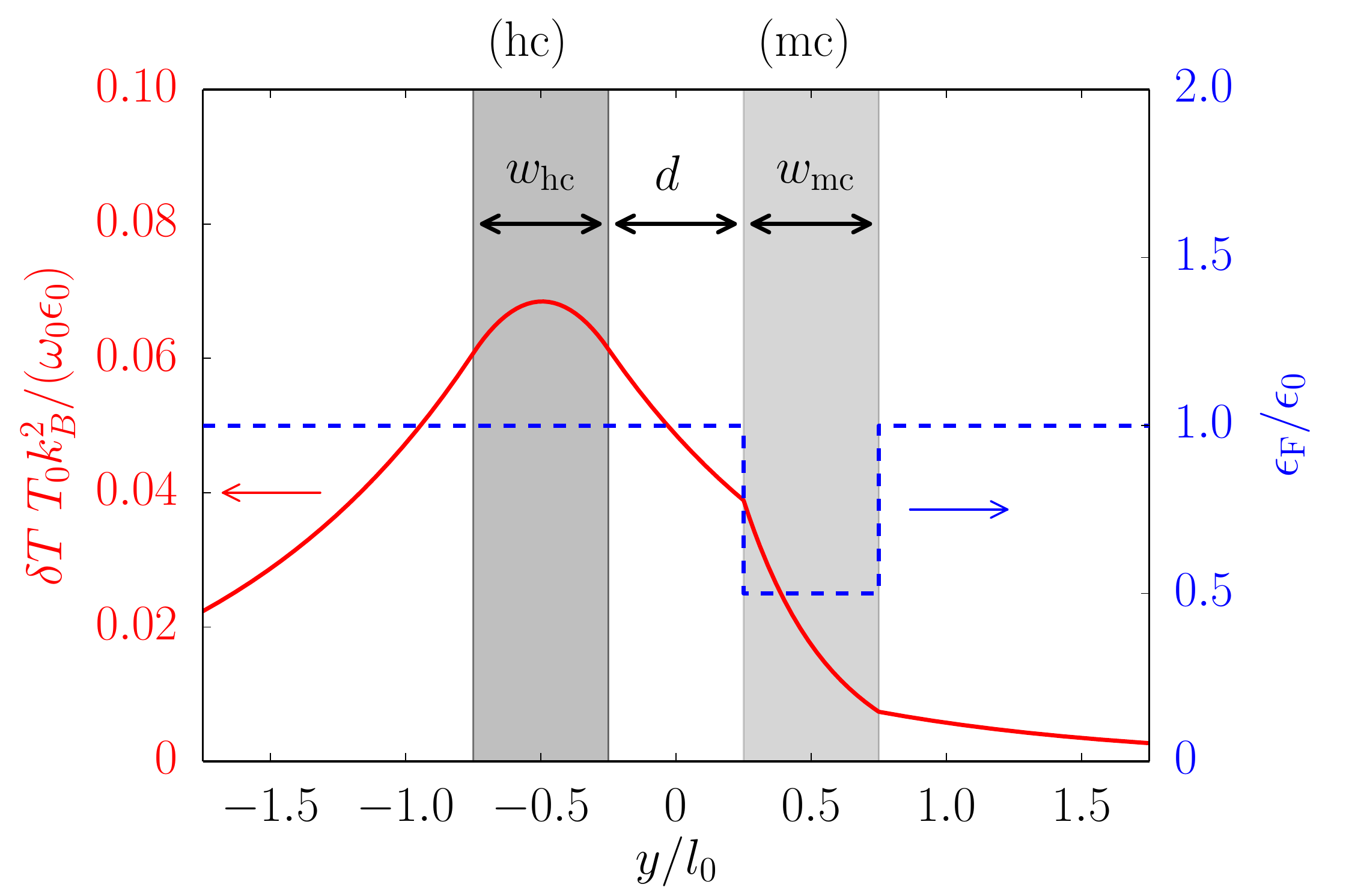}
	\caption{\label{TempPro} (Color online) Setup consisting of one heating (hc) and one modulation (mc) channel in a distance $d=l_0/2$ from the heated region. The width of both is $w_\text{hc}=w_\text{mc}=l_0/2$.  The red, solid line marks the temperature difference to the base temperature, $T_0$. The local Fermi energy is depicted by the blue, dashed line.}
\end{figure}

A generic setup for addressing mesoscopic diffusion thermopower is shown in Fig.~\ref{TempPro} consisting of one heating channel (hc) of width $w_\text{hc}$ and one modulation channel (mc) of width $w_\text{mc}$. 
The modulation channel is separated from the heating channel by a  distance $d$. 
Between the modulation and the heating channel, the carrier density remains unchanged and no electric field is applied, $E_x=0$.
The blue, dashed line (right $y$-axis) in Fig.~\ref{TempPro} depicts the profile of the Fermi energy.
The resulting temperature profile is marked by the red, solid line (left $y$-axis).

Within the Sommerfeld expansion (see Sec.~\ref{sommerfeld}), the values of $\tau_\text{e}$ and $\tau_\text{i}$ and their derivatives at the local Fermi energy are needed. We model the full energy dependence of the relaxation times\cite{Walukiewicz1984, Harris1989,Hwang2008a} by power laws $\tau_\textrm{e} = \tau_{\textrm{e},0} \left(\epsilon/\epsilon_0\right)^{\alpha_\textrm{e}}$ and $\tau_\textrm{i} = \tau_{\textrm{i},0} \left(\epsilon/\epsilon_0\right)^{\alpha_\textrm{i}}$.
The values $\tau_{\textrm{e},0}$ and $\tau_{\textrm{i},0}$ of the relaxation times at reference energy $\epsilon_0$ will, of course, enter the transverse voltage $U_y$.
By expressing all our results in terms of the proper dimensionless quantities, however, the values of $\tau_{\textrm{e},0}$ and $\tau_{\textrm{i},0}$ completely drop out.
They only need to be specified when comparing to experimental data.
For the exponents, we take $\alpha_\text{e} = 0.88$ and $\alpha_\text{i} = 1.45$, that were experimentally determined for the device used in Ref.~\onlinecite{ganczarczyk_transverse_2012}, but as we demonstrate below, a variation of $\alpha_\text{e/i}$ leads to only small quantitative corrections.

\begin{figure}
\includegraphics[width=0.49 \textwidth]{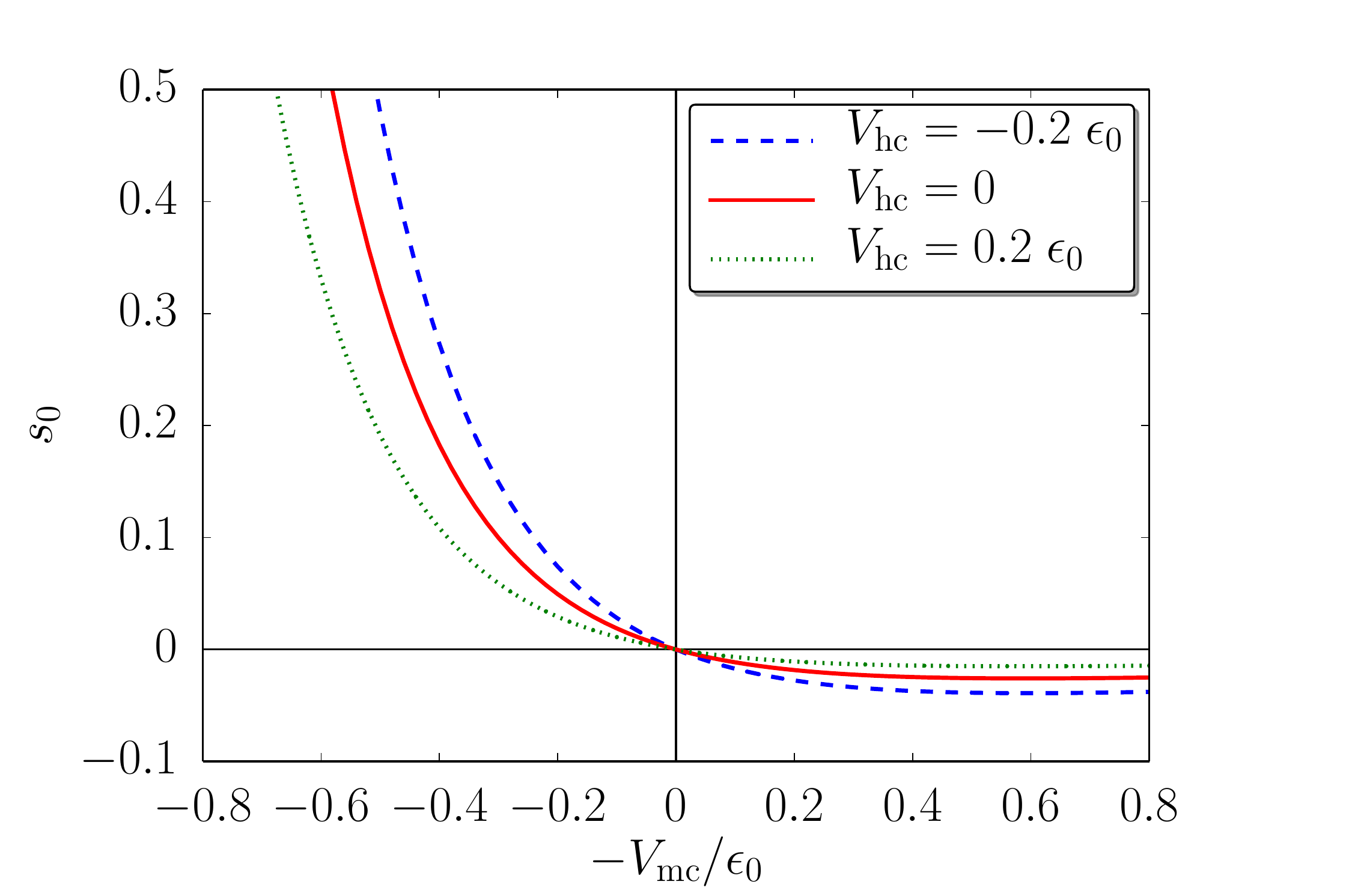}
\caption{\label{VR} (Color online) Dependence of $s_0$ on the electrostatic potential $V_\text{mc}$ in the modulation channel for three different values of the potential $V_\text{hc}$ in the heating channel. Referring to the setup in Fig.~\ref{TempPro}, we set $w_\text{hc}=w_\text{mc}=l_0/2$ and $d=0$.}
\end{figure}

In Fig.~\ref{VR} we show $s_0$ as a function of the potential $V_\text{mc}$ that the electrons experience in the modulation channel due to the external gate voltage. For clarity, we study the case of vanishing gap ($d=0$) between heating and modulation channel and a fixed channel width first. Those mesoscopic aspects are addressed in Sec.~\ref{meso}. The results shown in Fig.~\ref{VR} are calculated for the  setup in Fig.~\ref{TempPro} with $w_\text{hc}=w_\text{mc}=l_0/2$ and $d=0$. (The potential in the heating channel, $V_\text{hc}$, and the modulation channel, $V_\text{mc}$, are obtained from the applied gate voltage by multiplying with the electron charge $-e$ and a device-dependent lever factor.)  
For vanishing potential in the modulation channel, $V_\text{mc}=0$, the setup is spatially symmetric and the transverse voltage vanishes accordingly. 
A positive modulation potential $V_\text{mc}$ decreases the Fermi energy, $\epsilon_\text{F} = \epsilon_0-V_\text{mc}$, and, thus, the carrier density $n$ in the modulation channel, which leads to an enhanced Seebeck coefficient $S$ there. 
That results in a positive $s_0$, which diverges close to the depletion of the 2DEG in the modulation channel (for $V_\text{mc}\rightarrow \epsilon_0$). For an enhanced carrier density, on the other hand, $s_0$ becomes negative since the Seebeck coefficient in the modulation channel is reduced compared to the ungated region of the 2DEG on the opposite side of the heating channel.
The solid (red) line in Fig.~\ref{VR} depicts $s_0$ for the absence of a modulation of the carrier density within the heating channel.
For comparison, the dashed (blue) line and dotted (green) line in Fig.~\ref{VR} represent results with a finite potential within the heating channel. 
Since the applied voltage that drives the current in $x$-direction is kept constant, the density modulation in the heating channel influences the current and, thereby, the energy acquired by the electrons. That leads to a larger amplitude of $s_0$ for negative $V_\text{hc}$ and vise versa. 
For the rest of the paper (except when discussing smeared potential steps in Sec.~\ref{sharp}), we set $V_\text{hc}=0$, which yields $s=s_0$.

\begin{figure}
\includegraphics[width=0.49 \textwidth]{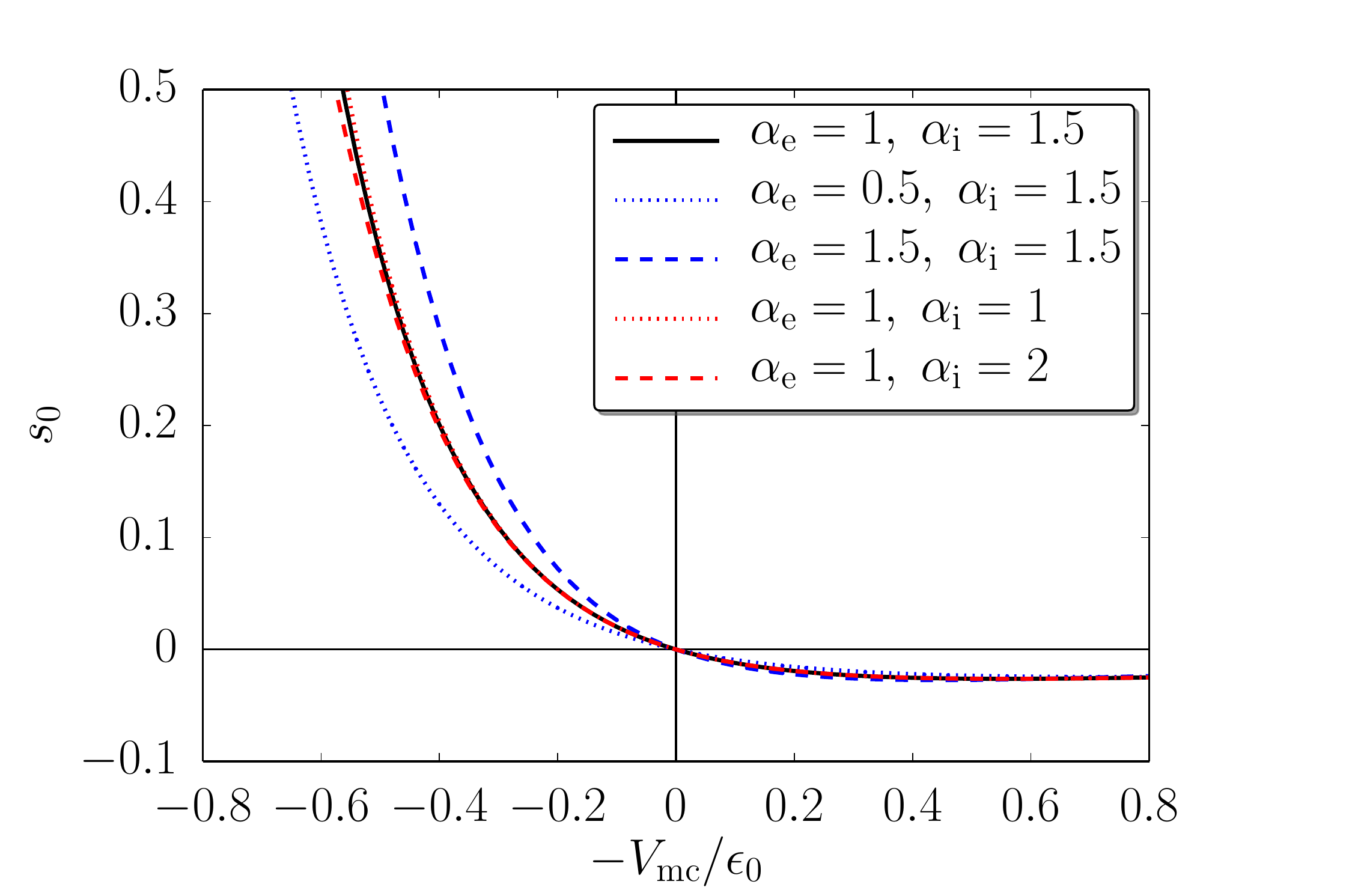}
\caption{\label{G}(Color online) Dependence of $s_0$ on the electrostatic potential in the modulation channel $V_\text{mc}$ comparing different $\alpha_\text{e}$ and $\alpha_\text{i}$. Referring to the setup in Fig.~\ref{TempPro}, we set $w_\text{hc}=w_\text{mc}=l_0/2$ and $d=0$.}
\end{figure}

In Fig.~\ref{G}, we compare the results for five different sets of exponents $\alpha_\text{e}$ and $\alpha_\text{i}$ (using again $w_\text{hc}=w_\text{mc}=l_0/2$ and $d=0$).
We find that the variation of $s_0$ is small even for the large range of chosen $\alpha_\text{e}$ (from $0.5$ to $1.5$) and $\alpha_\text{i}$ (from $1$ to $2$).
This shows that the specific values of the exponents are of minor importance.

\subsection{Multiple Heating Channels\label{MHC}}

Almost entirely throughout this paper, we consider setups with only one heating channel.
By this, we avoid structures that are experimentally more difficult to realize and control.
But, more importantly, the restriction to one heating channel only is motivated on the theoretical ground by the linear structure of the differential equation Eq.~\eqref{Tdiff} with respect to $\omega$.
If there are several heating channels, then the total temperature increase is just the sum of the individual temperature increases $\delta T_j(y) = T_j(y) -T_0$ that would arise if only heating channel $j$ was carrying an electric current while all other channels were kept current free.
Therefore, considering only one heating channel does not define a conceptual restriction.

The transverse rectifier studied in Ref.~\onlinecite{ganczarczyk_transverse_2012} is an example of a device with effectively two heating channels since electric current was passed through two connected, parallel 2DEG stripes with different carrier densities.
As stated above, the total output voltage is, in this case, the sum of the voltages generated by the heating currents in the individual channels.
Furthermore, this is an example of the case that heating and modulation channel coincide, while for most of the results shown in the present paper we assume the heating channel not to be gated.

\section{Mesoscopic Effects}
\label{meso}

In order to discuss mesoscopic aspects of the diffusion thermopower, we consider a suitable reference in the macroscopic regime first.  For this, we study an ungated device, $V_\text{hc}=V_\text{mc}=0$, with a wide heating channel, $w_\text{hc} \gg l_0$. Then, the voltage $U_y$ drop between a position deep inside the heating channel and a contact far outside leads to the result 
\begin{equation}
\label{s_macro}
	s_0 = 1 + \alpha_\text{e} \, , 
\end{equation}
which depends only on one, material-specific parameter $\alpha_\text{e}$.
Further details of calculation that lead to Eqs.~\eqref{s_macro}-\eqref{s_depleted} are presented in App.~\ref{App}. 

Device geometries as depicted in Fig.~\ref{fig:setup}, however, in which the transverse voltage is measured spatially separated from the heating channel, can not access the individual voltage drop between heating channel and left or right contact. 
Instead, the difference between the contributions from the left and the right hand side is measured, which immediately yields $s_0=0$ in case of a symmetric device.
Here, the modulation channel with $V_\text{mc}\neq 0$ on the right side of the heating channel breaks the left/right symmetry.
To remain in the macroscopic limit, we take $w_\text{mc} \gg l$ and set $d=0$.
If the electron density in the modulation channel is reduced, the Seebeck coefficient is increased there, i.e., the contribution from the right part of the device to the thermopower dominates over the one from the left, so that $s_0>0$ (see Fig.~\ref{VR}). 
We remark that the contribution from the modulation channel to the total transverse voltage, $S \Delta T$, depends on the temperature difference $\Delta T=T_\text{i}-T_0$ between the temperature $T_\text{i}$ at the heating-/modulation-channel interface and the base temperature $T_0$.
The interface temperature, $T_\text{i}$, which differs from $T_\text{bulk}$ reached deep inside the heating channel, depends on $V_\text{mc}$ via
(in the macroscopic regime)
\begin{align}
	\label{Tinterface}
	\frac{T_\text{i}-T_0}{T_\text{bulk}-T_0} = \frac{1}{1+(1-V_\text{mc}/\epsilon_0)^{(1+\alpha_\text{e}-\alpha_\text{i})/2}} \, .
\end{align}
For $V_\text{mc} = 0$, the interface temperature is just the arithmetic mean $T_\text{i} = (T_\text{bulk}+T_0)/2$, and in the limit of an almost depleted modulation channel, $\epsilon_0-V_\text{mc}\ll \epsilon_0$, the interface temperature $T_\text{i} $ approaches $T_\text{bulk}$.

The $V_\text{mc}$ dependence of the Seebeck coefficient in the modulation channel is given by $\epsilon_0/(\epsilon_0-V_\text{mc})$, i.e., it becomes large in the limit of an almost depleted modulation channel.
In this case, the contributions to the transverse voltage outside the modulation channel can be neglected.
Setting, furthermore, $T_\text{i} \approx T_\text{bulk}$, we find that 
\begin{equation}
\label{s_depleted}
	s_0 \approx \frac{1 + \alpha_\text{e}}{1-V_\text{mc}/\epsilon_0} 
\end{equation}
diverges when approaching full depletion.

In the mesoscopic regime, the measured diffusion thermopower depends on the geometric lengths of the device. 
First, once the width of the heating channel becomes comparable to the energy diffusion length, $w_\text{hc} \sim l_0$, the maximum of the temperature profile within the heating channel is substantially smaller than the value $T_\text{bulk}$ reached in the macroscopic limit. 
Thus, the interface temperature $T_\text{i}$ is smaller and the transverse voltage, too. 
Second, a modulation-channel width comparable to the energy diffusion length, $w_\text{mc} \sim l$, has the effect that the temperature cannot fully drop to the base temperature $T_0$ in the modulation channel.
This also leads to a reduction of $s_0$.
Third, the energy diffusion also defines the length scale on which heating and modulation channel can be separated without loosing the transverse voltage: as long as $d \lesssim l_0$, the electrons in the modulation channel will experience a temperature gradient and $s_0$ will be finite.
A fourth mesoscopic effect is associated with the variation of the charge-carrier density on a length $\lambda$ for smeared potential steps.

\subsection{Heating-Channel Width}

\begin{figure}
\includegraphics[width=0.49 \textwidth]{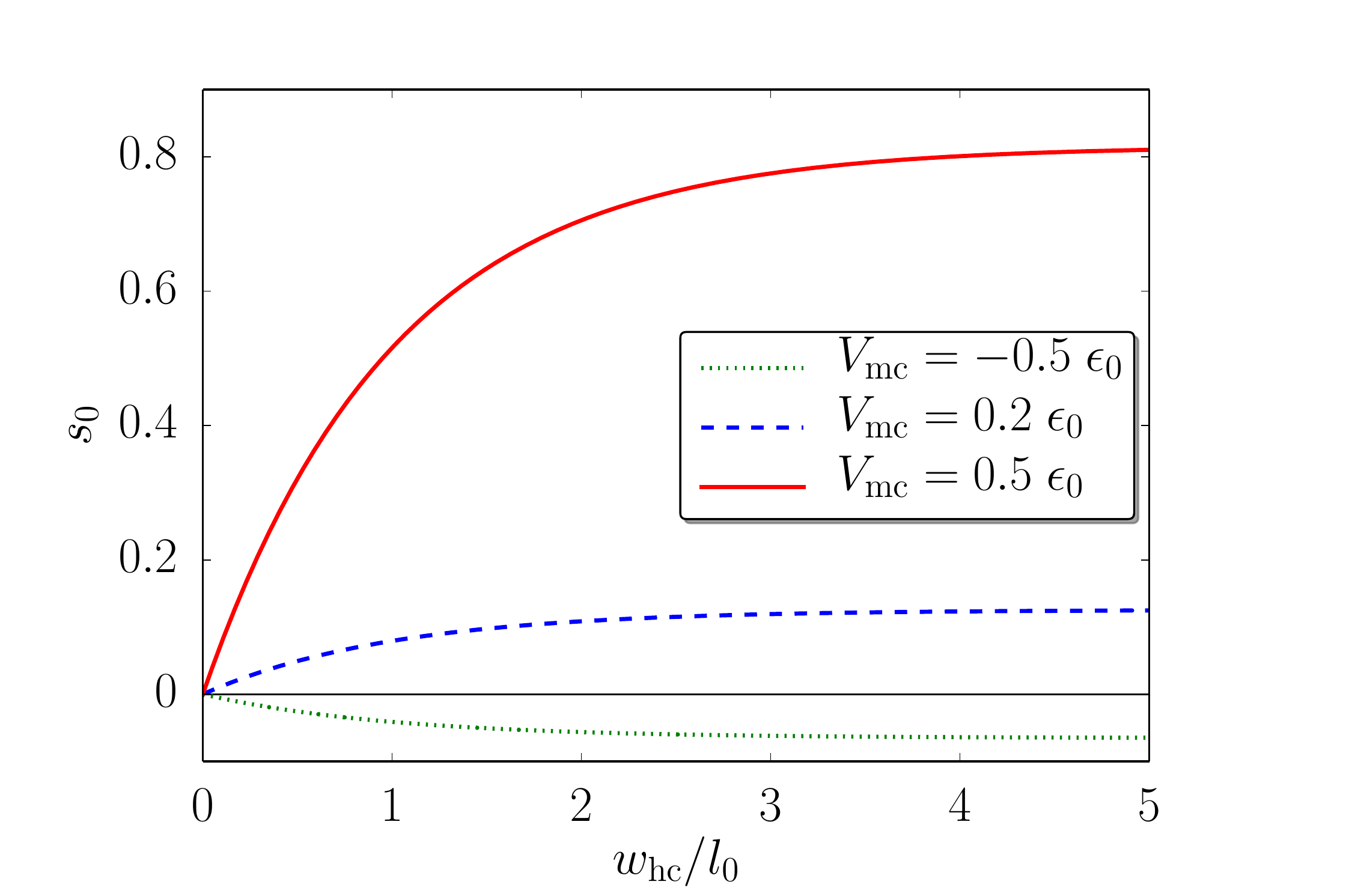}
\caption{\label{WL} (Color online) Dependence of $s_0$ on the width of the heating channel $w_\text{hc}$ for the potential $V_\text{mc}$ in the modulation channel. Referring to the setup in Fig.~\ref{TempPro}, we set $w_\text{mc}=l_0/2$ and $d=0$.}
\end{figure}

First, we discuss the role of the heating-channel width.
For this, we consider a device with $d=0$ (no space between heating and modulation channel) and fixed potential $V_\text{mc}$. As we stated  before, we restrict the results presented here to the case $V_\text{hc}=0$. 
The energy diffusion length sets the scale of spatial temperature variations. Explicitly, the vanishing potential in the heating channel fixes the energy diffusion length there.
As an important consequence, the maximal possible value $T_\text{bulk}$ of the temperature in the heating channel can only be asymptotically reached if the heating-channel width $w_\text{hc}$ is large compared to the energy diffusion length.
In the opposite limit, the maximal temperature in the heating channel is much lower than $T_\text{bulk}$. 
This means that the temperature drop across the modulation channel and, therefore, also the transverse voltage is reduced.
This reasoning is consistent with the results shown in Fig.~\ref{WL}, where the dependence of $s_0$ on $w_\text{hc}$ is depicted for $w_\text{mc}=l_0/2$ (and $d=0$): for $w_\text{hc} \lesssim l_0$, the transverse voltage and, thus, $s_0$ is reduced. In case of vanishing $V_\text{hc}$, the functional dependence of $s_0$  on $w_\text{hc}$ is exactly given by the exponential relation $s_0(w_\text{hc})= s_0(\infty) \left[1- \exp(-w_\text{hc}/l_0)\right]$. 

The heating-channel width should be compared with the energy-diffusion length of the heating channel.
The latter could be tuned via a nonvanishing $V_\text{hc}$.
For a positive $V_\text{hc}$, the energy diffusion length is reduced such that the suppression of $U_y$ is less severe for a fixed value of $w_\text{hc}$.
However, also the energy $\omega$ decreases, and this has a much stronger effect on $s_0$ than the reduction of $l$.

\subsection{Modulation-Channel Width}

\begin{figure}
\includegraphics[width=0.49 \textwidth]{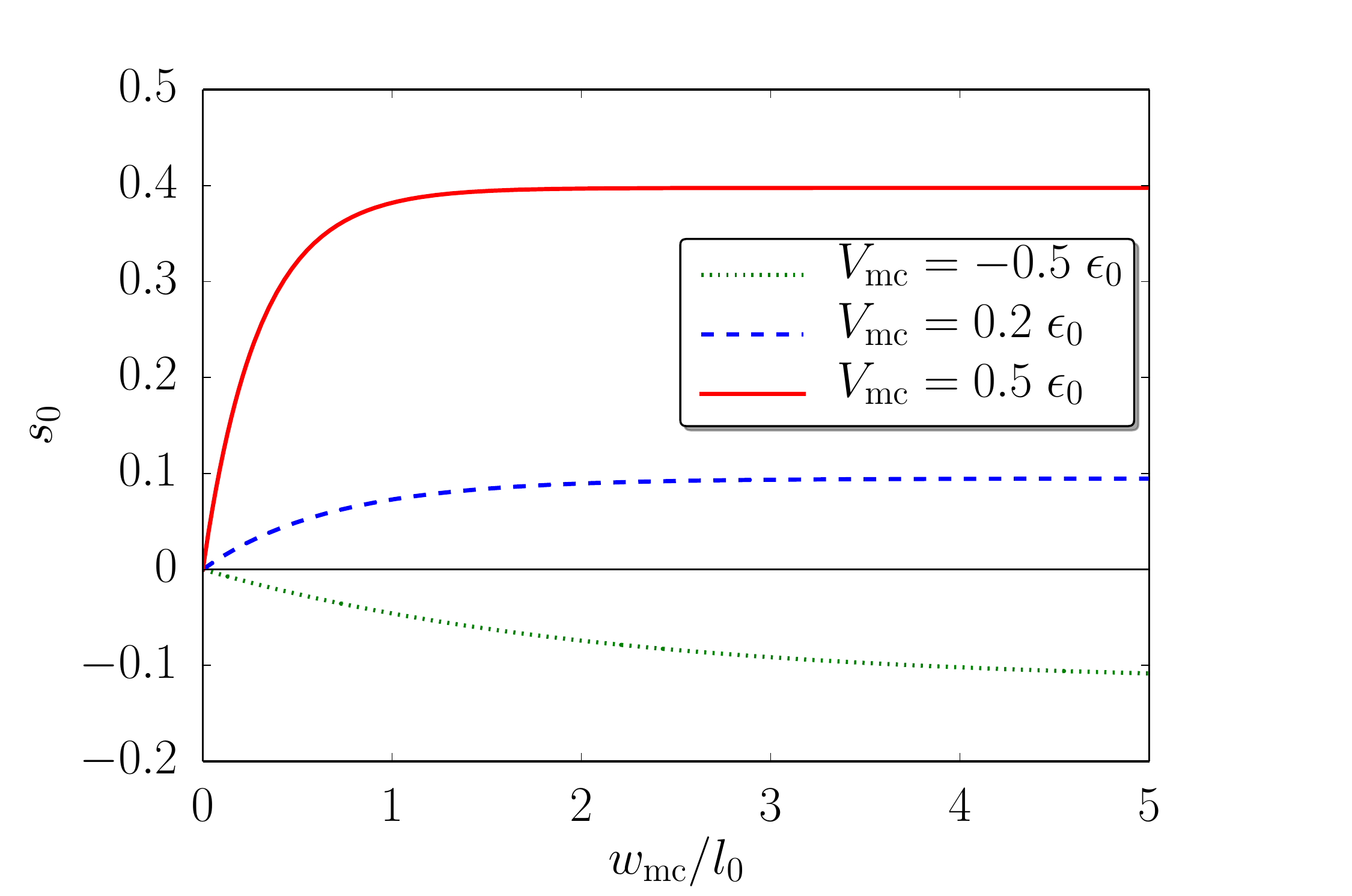}
\caption{\label{WR} (Color online) Dependence of $s_0$ on the width of the modulation channel $w_\text{mc}$ for the potential $V_\text{mc}$ in the modulation channel. Referring to the setup in Fig.~\ref{TempPro}, we set $w_\text{hc}=l_0/2$ and $d=0$.}
\end{figure}

Next, we discuss the dependence of $s_0$ on the modulation-channel width $w_\text{mc}$ while keeping the heating-channel width fixed.

Again, we take $d=0$ (no space between heating and modulation channel) and $V_\text{hc}=0$. The finite potential $V_\text{mc}$ in the modulation channel fixes the energy diffusion length there.  
In Fig.~\ref{WR}, we show the $w_\text{mc}$ dependence of $s_0$ for $w_\text{hc}=l_0/2$. 
We, again, find an exponential behavior.

The relevant length scale is the energy diffusion length, $l$, in the modulation channel.
For $w_\text{mc} \gg l$, the temperature profile is such that the temperature at the right edge of the modulation channel already reaches the base temperature $T_0$.
A large temperature drop within the modulation channel is accompanied by a large value of $s_0$.
For $w_\text{mc} \lesssim l$, this temperature drop is reduced and $s_0$ is smaller.

The fact that the energy diffusion length $l$ in the modulation channel depends on the potential $V_\text{mc}$ can be clearly seen in Fig.~\ref{WR}.
For positive values of $V_\text{mc}$ (decreased electron density) the energy diffusion length $l$ becomes smaller than the reference $l_0$, for negative values of $V_\text{mc}$, $l$ is larger.

\subsection{Heating-/Modulation-Channel Distance}

\begin{figure}
\includegraphics[width=0.49 \textwidth]{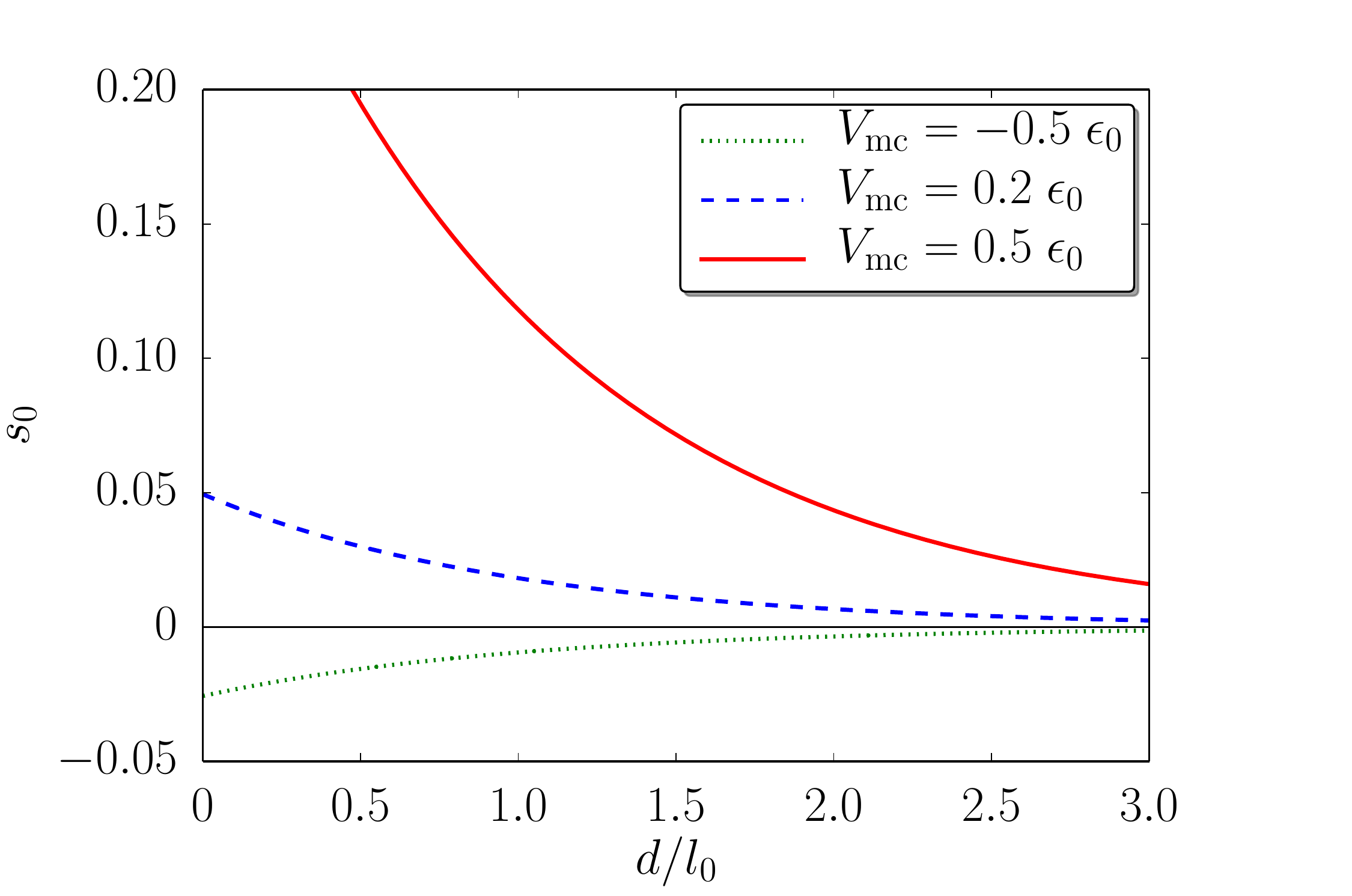}
\caption{\label{Gap} (Color online) Dependence of $s_0$ on the distance $d$ between heating and modulation channel for different potentials $V_\text{mc}$ in the modulation channel. Referring to the setup in Fig.~\ref{TempPro}, we set $w_\text{hc}=w_\text{mc}=l_0/2$.}
\end{figure}

Not only the widths of heating and modulation channel matter, but also their separation $d$ from each other.
For separations $d$ much larger than the energy diffusion length $l_0$, the elevated temperature in the heating channel drops down to base temperature before reaching the beginning of the modulation channel.
As a consequence, there is no temperature drop across the modulation channel, and the transverse voltage vanishes.
A finite $s_0$ requires $d\lesssim l_0$.
This is illustrated in Fig.~\ref{Gap} for different values of $V_\text{mc}$.

\subsection{Sharpness of Potential Step \label{sharp}}

\begin{figure}
	\includegraphics[width=.49\textwidth]{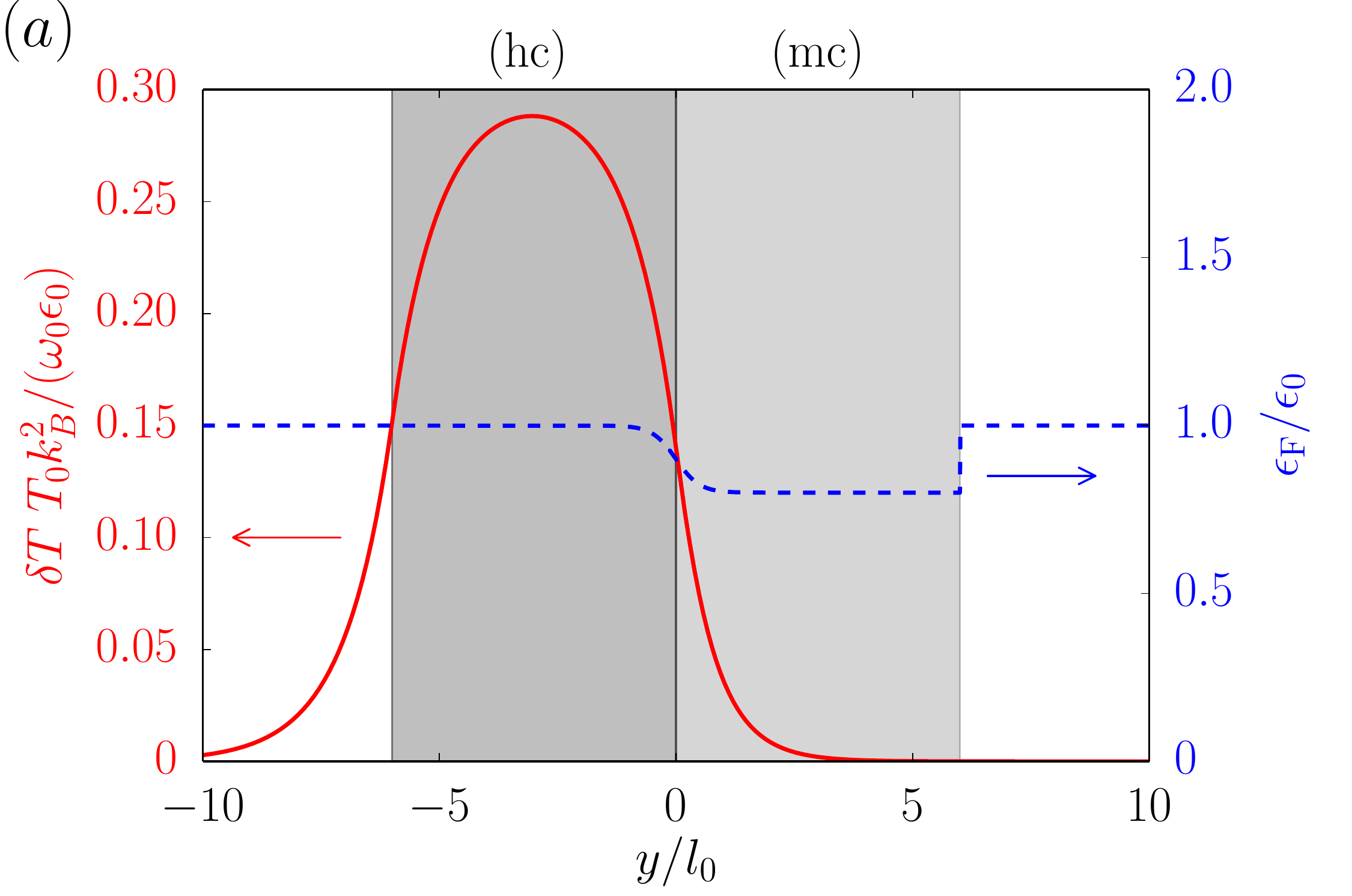}
	\includegraphics[width=.49\textwidth]{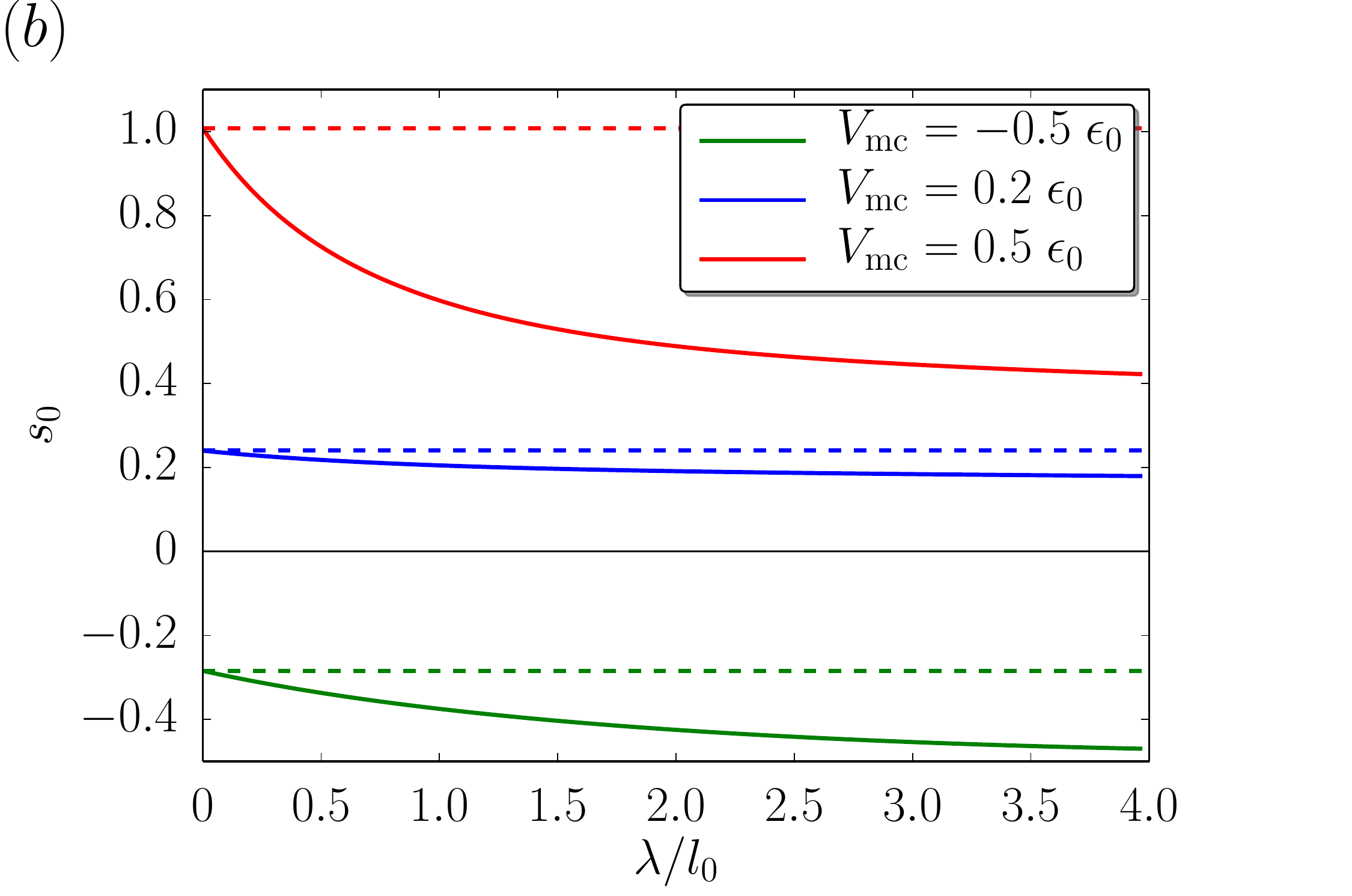}
		\caption{\label{Step}(Color online) Dependence of $s_0$ on the sharpness of the potential step between modulation and heating channel. The width of the channels is set to $w_\text{hc}=w_\text{mc}=6\, l_0$ and the distance $d=0$. (a) The step of the Fermi energy at the heating-/modulation-channel interface is given by $\epsilon_\text{F}(y)=\epsilon_0-V(y)$ in the range $-6\, l_0<y<6\, l_0$ shown by the blue dashed line. Here, the potential $V(y)$ is taken from Eq.~\eqref{V}. The red line marks the temperature difference, $\delta T(y)$, to the base temperature, $T_0$. For (a) we use $V_\text{mc}=0.2\, \epsilon_0$  and $\lambda=l_0/2$. (b) The dashed lines mark $s_0$ for a sharp step potential for different potentials $V_\text{mc}$ in the modulation channel. The solid lines depict the influence of the step width $\lambda$. }
\end{figure}

Another length that may be relevant in some devices is the width on which a potential step is smeared.
So far, we have always assumed sharp steps.
To investigate the role of a finite step width, we construct a smooth potential profile which changes on a characteristic length $\lambda$. 
As an example, we take
\begin{align}
	V(y)=V_\text{mc} \frac{\tanh\left(2y/\lambda\right)+1}{2}\label{V}
\end{align}
depicted in panel (a) of Fig.~\ref{Step}.
Here, $V_\text{mc}$ is the height and $\lambda$ the width of the step.
The modulation and heating channel are kept at a constant width of $w_\text{hc}=w_\text{mc}=6\, l_0$, which allows for a variation of $\lambda$ on an appropriate range without side effects.

Figure~\ref{Step}~(b) shows the dependence of $s_0$ on $\lambda$ for different step heights $V_\text{mc}$. 
Again, we find an exponential dependence of $s_0$ on a length scale given by the energy diffusion length. 
The finiteness of the step width influences $s_0$ in different ways. 
First, the energy diffusion length $l$ is changed over a region of length $\lambda$ across the interface, which, in turn, influences the temperature profile. 
Second, the Seebeck coefficient $S$ becomes $y$-dependent.
Third, the amount of excess energy $\omega$ provided by the heating current is modified at the edge of the heating channel. 
The combination of the three gives rise to the behavior displayed in Fig.~\ref{Step}~(b).
For negative values of $V_\text{mc}$, the effect on $\omega$ dominates and the amplitude of $s_0$ gets larger with increasing $\lambda$.
For positive values of $V_\text{mc}$, on the other hand, the effects on $S$ and $l$ are more important and $s_0$ decreases as function of $\lambda$.

\subsection{Additivity}

In order to amplify the output transverse voltage, one may want to put $n$ devices (each consisting of one heating and one modulation channel) in series (similarly as it is done for electron ratchets\cite{blanter,Sassine2008,Olbrich2011} or in the context of state-dependent diffusion \cite{buettiker}).
Then the important question arises, whether the $n$ devices simply add, i.e., whether the total output voltage just $n$ times the voltage of a single device.

\begin{figure}
\includegraphics[width=0.49 \textwidth]{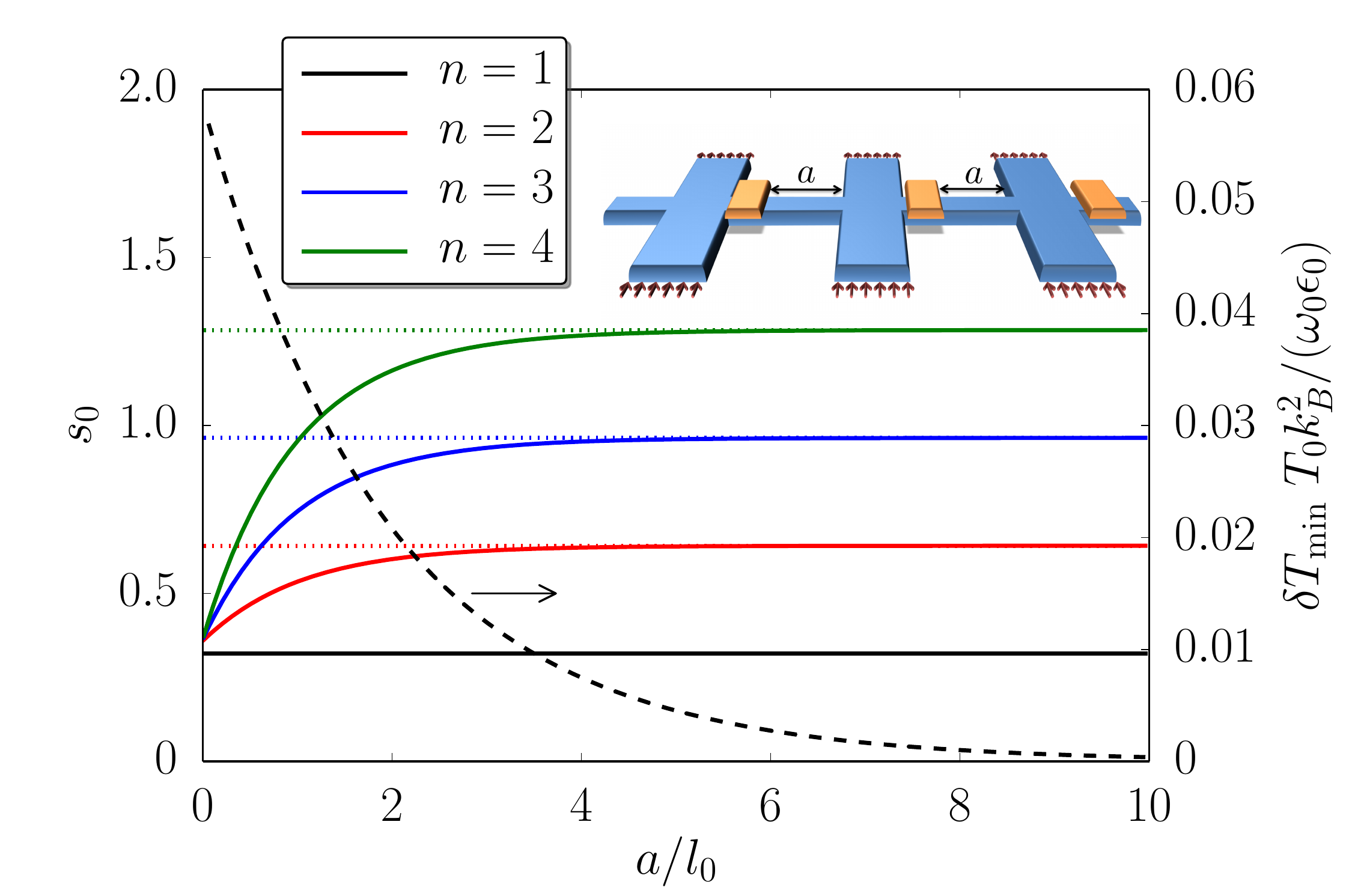}
\caption{\label{Add}(Color online)  The solid lines mark $s_0$ for $n$ devices as function of $a$. The black, dashed line show the minimum temperature value between the devices, also as function of $a$. The inset depicts an example consisting of three devices (with $w_\text{hc}=w_\text{mc}=l_0/2$ each), separated by a distance $a$.}
\end{figure} 

To answer this question, we model a structure as shown in the inset of Fig.~\ref{Add}.
For each single device we choose $w_\text{hc}=w_\text{mc}=l_0/2$ and $d=0$.
Neighboring devices are separated by an ungated and current-free channel of the length $a$.
The applied electrical field $E_x$ driving the heating currents is the same in all devices.
As discussed in Sec.~\ref{MHC}, there is a superposition principle for the heating currents: the profile of the total temperature increase is just the sum of the temperature-increase profile $\delta T_j$ due to device $j=1, \ldots, n$.
We find that for $a \gg l_0$, the total $s_0$ is just $n$ times the result for a single device, i.e., additivity holds.
Neighboring devices do not influence each other since the increased temperature due to heating in device $j$ has already dropped down to base temperature $T_0$ before reaching the neighboring devices $j\pm1$ (see dashed line in Fig.~\ref{Add} which shows the minimum value $\delta T_\text{min}$ of the temperature increase in between neighboring devices).  
This is different for $a \lesssim l_0$.
In this case, the temperature profiles generated by neighboring devices influence each other, and the overall performance is reduced.
In the extreme limit $a=0$, the total performance is independent of the number $n$ of devices.
This can be easily understood by observing that the devices in the middle build a symmetric potential landscape of the modulation channels such that only the edge devices contribute, which is equivalent to $n=1$.

\subsection{Relation to Recent Experiments}

The output voltage measured in Ref.~\onlinecite{ganczarczyk_transverse_2012} as function of two gate voltages could only be explained by including the mesoscopic effects due to finite heating- and modulation-channel widths as compared to the energy diffusion length.  
This was, actually, the motivation for the systematic study of the mesoscopic aspects presented here.

But there is also an earlier measurement of the diffusion thermopower in a 2DEG realized in a GaAs/AlGaAs heterostructure,\cite{Chickering2009} in which mesoscopic effects due to the finite modulation-channel width may play a role. 
Chickering \textit{et al.} found a dependence of the thermopower on temperature and electron density that is compatible with the Mott formula.

For temperatures $T_0\lesssim2\, \text{K}$, the absolute value of the thermopower was, however, reduced by about $20\%$  as compared to what they expected in their analysis. Thereby, they did not take mesoscopic effects into account.

\begin{figure}
	\includegraphics[width=.49\textwidth]{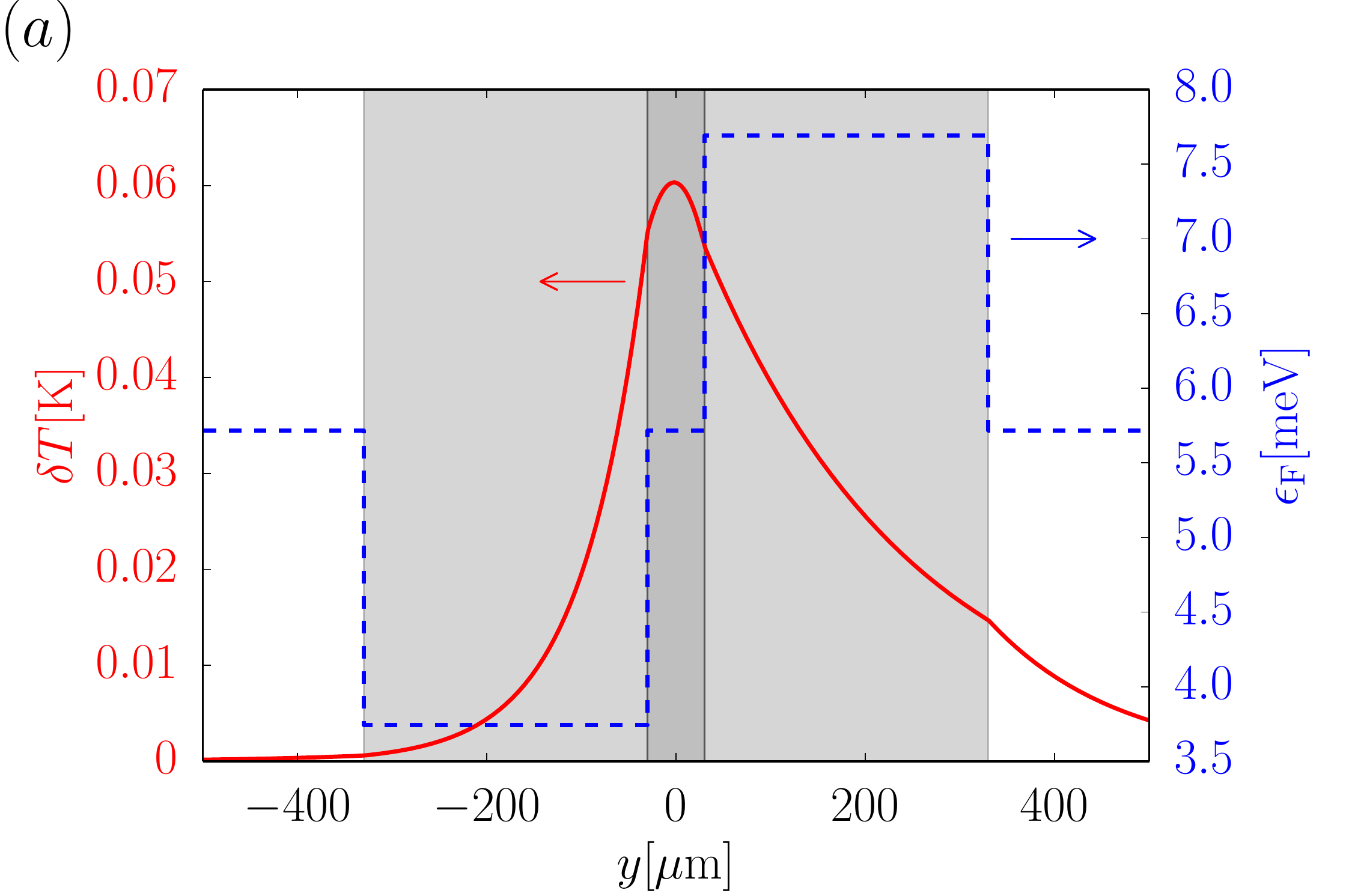}
	\includegraphics[width=.49\textwidth]{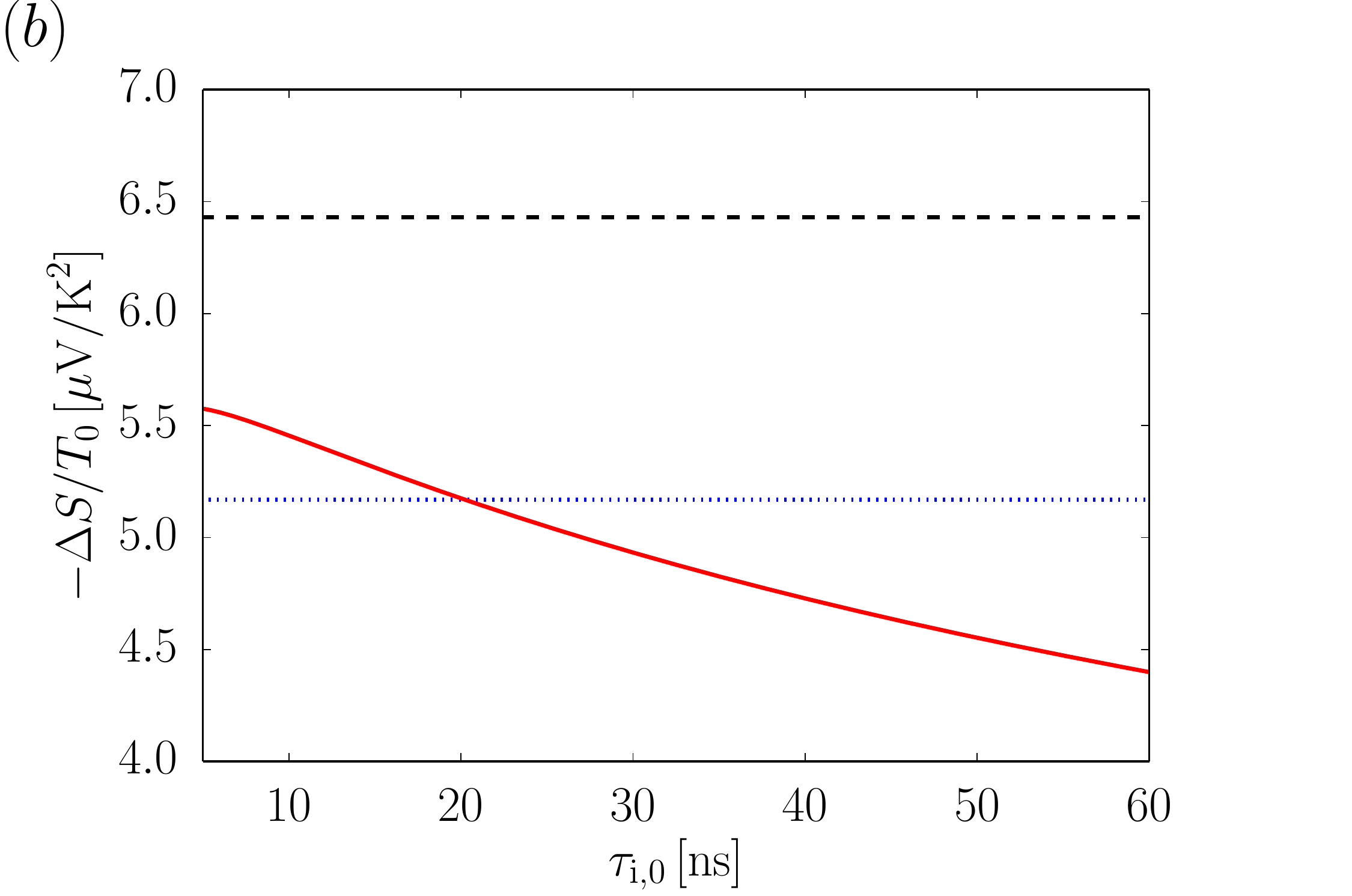}
		\caption{\label{fig:Eisen}(Color online) Results for a setup consisting of one heating channel of width $w_\text{hc}=60\, \mu\text{m}$ and two neighboring modulation channels of widths $w_\text{hc}=300\,  \mu\text{m}$ each. The calculations are done for $T_0=2\, \text{K}$ and the momentum relaxation time $\tau_{\text{e},0}=0.126\,  \text{ns}$ (extracted from the mobility in Ref.~\onlinecite{Chickering2009}).
		(a)~The red, solid line marks the profile of $\delta T$ and the local Fermi energy is depicted by the blue, dashed line.
		We choose $\tau_\text{i,0}=100\,    \tau_\text{e,0}$ for the energy-relaxation time. 
		(b)~The black, dashed and the blue, dotted lines represent the theoretical and measured results for thermopower divided by the lattice temperature obtained by Chickering~\textit{et~al.}, respectively.\cite{Chickering2009}
		The red, solid line accounts for mesoscopic effects.}
\end{figure}

We simulated the device used in Ref.~\onlinecite{Chickering2009}.
The potential profile that accounts for the data in Fig.~2 of Ref.~\onlinecite{Chickering2009} for $\Delta(1/n)=4.9\times 10^{-12}\,  \text{cm}^2$ at $2\, \text{K}$ is shown in Fig.~\ref{fig:Eisen}~(a). Here, $\Delta(1/n)$ is the difference of the reciprocal carrier density between the right and left modulation channel.
The dark gray region in Fig.~\ref{fig:Eisen}~(a) indicates the narrow heating channel of width $w_\text{hc}=60\, \mu\text{m}$.
Two modulation channels (light gray) of width $w_\text{mc}=300\, \mu\text{m}$ are placed left and right of the heating channel.
We are able to identify the momentum-relaxation time $\tau_\text{e,0}= 0.126\, \text{ns}$ from the electric conductivity under the assumption that the energy-relaxation time $\tau_\text{i,0}$ is much larger. In this case we can set the exponent $\alpha_\text{e}$ to the corresponding value, $\alpha_\text{e}=0.9$, used by Chickering~\textit{et~al.}.
Unfortunately, the value for $\tau_\text{i,0}$ and $\alpha_\text{i}$ are not known for this experiment.
For the exponent $\alpha_\text{i}$, we take the same value as determined in Ref.~\onlinecite{ganczarczyk_transverse_2012}. As discussed previously, the effects of that parameter are of minor importance.

In panel (a) of Fig.~\ref{fig:Eisen}, we show the calculated temperature profile if we assume {$\tau_\text{i,0}= 100\, \tau_\text{e,0}$, which is close to the ratio found for the device in Ref.~\onlinecite{ganczarczyk_transverse_2012}. 
The fact that the temperatures on the outer edges of the modulation channels have not yet reached base temperature indicates that mesoscopic corrections due to a finite modulation-channel width have to be expected.
Mesoscopic effects due to the finite heating-channel, on the other hand, are not relevant for the $20\%$ deviation of the measured thermopower since Chickering~\textit{et~al.} measured the average temperature in the heating channel instead of calculating $T_\text{bulk}$ from the heating current.
In panel (b) of Fig.~\ref{fig:Eisen}, we compare the results of Ref.~\onlinecite{Chickering2009} for  $\Delta(1/n)=4.9\times 10^{-12}\, \text{cm}^2$ at $2\,\text{K}$ to our simulation that includes mesoscopic effects. That is done in dependence of the unknown quantity $\tau_\text{i,0}$  which crucially influences the mesoscopic effects since the energy relaxation time enters the energy diffusion length. The black, dashed line in Fig.~\ref{fig:Eisen}~(b) corresponds to the result determined by the Mott formula \eqref{thermopower}. The difference of the thermopower in the right and left modulation channel, $\Delta S$, is divided by the base temperature $T_0$. The blue, dotted line shows the measured $-\Delta S/T_0$, where the amplitude is reduced by about $20\%$ compared to the dashed line. To include mesoscopic aspects, we calculate the transverse voltage and divide it by both $T_0$ and the average (measured) temperature in the heating channel which yields the red, solid line in Fig.~\ref{fig:Eisen}~(b).
We find a reduction of about $15\%$ to $25\%$ for realistic choices of the energy-relaxation time.

\section{Energy Diffusion Length}

For studying thermoelectric effects which are associated with the electron temperature and energy diffusion in 2DEGs, it is important to determine the energy diffusion length.
To do so in an indirect way, one may measure the momentum- and the energy-relaxation times $\tau_\text{e}$ and $\tau_\text{i}$.
This may be an easy task for $\tau_\text{e}$ (since it is related to the electrical conductivity), but determining $\tau_\text{i}$ is more challenging.
Our analysis, however, suggests that transverse thermoelectric rectifiers realized in 2DEGs are ideal systems to directly access the energy diffusion length by systematically varying the width of heating or modulation channels or distances between them: 
The dependence of $s$ (the mesoscopic analogon to the thermopower) on these lengths, calculated within the diffusion thermopower model, can be nicely approximated by exponential functions that depend on these lengths divided by the corresponding energy diffusion length $l$.

From the relation $l= v_\text{F}\sqrt{\tau_\text{e}\tau_\text{i}/2}$ it is immediately clear that the energy diffusion length can be tuned by applied gate voltages via the energy dependence of the Fermi velocity and the relaxation times.
Assuming that the energy dependences of $v_\text{F}$ and $\tau_\text{e}$ are known, a systematic variation of the gate voltage fixing the potential in a modulation channel of a transverse thermoelectric rectifier allows for an investigation of the energy dependence of $\tau_\text{i}$ which may be more difficult to access by alternative methods.

\section{Magnitude of Output Voltage}

The magnitude of the electric voltage generated by a given temperature difference is characterized by the Seebeck coefficient $S$.
For a 2DEG at low temperature, $S$ scales linearly with temperature and is proportional to the inverse of the electron density, $S \sim T/n$.
To characterize the performance of the all-electric devices studied in this paper, however, we use the dimensionless quantity $s=eU_y/\omega$.
Instead of relating the output thermoelectric voltage, $U_y$, to a temperature difference, which is not directly externally controlled but only appears indirectly as a consequence of electrical heating, we take as a reference the energy $\omega$ that is acquired per electron due to Joule heating.
As can be seen from Eq.~(\ref{s_macro}), $s$ is a dimensionless quantity of order one that does not scale with temperature $T$ or electron density $n$.
This can also be understood in the following way: 
Since the electrons form a Fermi gas, the extra energy $\omega$ deposited on average per particle by Joule heating is not uniformly distributed to all electrons but only to some fraction $k_\text{B}T/\epsilon_\text{F} = \rho k_\text{B}T/n$ around the Fermi level.
Therefore, the temperature increase due to Joule heating scales with $\omega n/T$ (see also Eq.~(\ref{tbulk})), which compensates the $T/n$ dependence of the Seebeck coefficient such that $s$ is of order one.

The input parameter in experiments is not $\omega$ but the bias voltage or, equivalently, the electric field $E_x$ driving the electric current which heats up the electrons. 
It may, therefore, be interesting to discuss the temperature and electron density dependence of $\omega$ for fixed $E_x$.
As can be seen from Eq.~(\ref{om_gen}) or (\ref{om_som}), $\omega$ scales with $n^{\alpha_\text{e}+\alpha_\text{i}}$, independent of temperature.

How can the output voltage be maximized for given input voltage?
The above mentioned compensation of the $T/n$ dependence of the Seebeck coefficient with the $\omega n/T$ dependence of the temperature increase due to Joule heating can be modified by choosing different electron densities $n_\text{hc}$ and $n_\text{mc}$ in the heating channel (responsible for the temperature increase) and the modulation channel (important for the thermoelectric voltage), respectively.
For a depleted modulation channel, $s$ is increased by a factor of $n_\text{hc}/n_\text{mc}$, see also Eq.~(\ref{s_depleted}) as well as Figs.~\ref{VR} and \ref{G}.

An important message of Sec.~\ref{meso} is that mesoscopic effects tend to reduce the output voltage.
To prevent this reduction, the width of both the heating and the modulation should be larger than the energy diffusion length $l$ see Figs.~\ref{WL} and \ref{WR}.
The distance between heating and modulation channel should, on the other hand, be smaller than $l$, see Fig.~\ref{Gap}.
And finally, for a series of multiple elements in the device, the distance between neighboring elements should be larger than $l$, see Fig.~\ref{Add}.

Furthermore, a series of elements as shown in Fig.~\ref{Add} provide the possibility for increasing the output voltage because the output signal scales with the number of elements (if the distance is larger than $l$). That concept of additivity is certainly a tool for optimization. 
However, it requires the same number of heating channels and, thus, the total input current increases simultaneously for fixed input voltage.

\section{Conclusions}

The characteristic length scales that determine the properties of charge transport and energy diffusion by electrons in 2DEGs strongly differ from each other at low temperature.
As compared to the elastic mean-free path $l_\text{e}= v_\text{F}\tau_\text{e}$ (that marks the crossover from ballistic to diffusive charge transport), the energy diffusion length $l= v_\text{F}\sqrt{\tau_\text{e}\tau_\text{i}/2}$ may be substantially larger.
The latter defines the scale for spatial variations of the local electron temperature in nonequilibrium situations evoked by local Joule heating.
On length scales comparable to this energy diffusion length, mesoscopic effects become important.
This gives rise to mesoscopic features of the diffusion thermopower.
Both the temperature profile and the value of electric voltage induced by local Joule heating depend on geometric dimensions of a specific device, including the width of heating channels, the width of modulation channels, or the separation between the two. 
It is, therefore, crucial that the theoretical modeling of thermoelectric devices realized in 2DEGs accounts for these mesoscopic effects. 
This can be conveniently done within a diffusion thermopower model as described in this paper.

\section*{Acknowledgements}

We acknowledge helpful discussions with James Eisenstein, Arkadius Ganczarczyk, Alfred Hucht, Ulrich Kunze, and Axel Lorke.
 Furthermore, we thank James Eisenstein for sending us an image of the device used in Ref.~\onlinecite{Chickering2009} from which we determined the widths of the heating and modulation channels.

\appendix

\section{Derivation of Eqs.~(\ref{s_macro}) - (\ref{s_depleted})\label{App}}

To derive Eq.~(\ref{s_macro}), we first determine the temperature increase deep inside the heating channel.
For this we use Eq.~(\ref{tbulk}) together with Eq.~(\ref{c_V}) to get 
\begin{equation}
	T_\text{bulk} - T_0 = \frac{3\omega \epsilon_0}{\pi^2 k_\text{B}^2T_0} \, .
\label{A1}
\end{equation}
Since the Seebeck coefficient
\begin{equation}
\label{A2}
	S = - \frac{\pi^2}{3} \frac{k_\text{B}^2T_0}{e} \frac{1+\alpha_\text{e}}{\epsilon_0}
\end{equation}
is constant in space, the integration in Eq.~(\ref{MottInt}) is trivial, which immediately yields $U_y = \omega/e$ and, as a consequence, Eq.~(\ref{s_macro}).

To determine the temperature $T_\text{i}$ at the interface at $y=0$ between an infinitely wide heating channel and and infinitely wide modulation channel at finite potential $V_\text{mc}$, we need to solve the differential equation Eq.~(\ref{Tdiff}) for $y<0$ and $y>0$ separately,
\begin{equation}
	T(y) = \left\{ 
	\begin{array}{cc}
	T_\text{bulk} + (T_\text{i}-T_\text{bulk}) \exp ( y/l_0) & \,\, \text{for} \,\,  y<0
	\\
	T_0+(T_\text{i} -T_0) \exp ( -y/l) & \,\, \text{for} \,\,  y>0
	\end{array}
	\right.
\end{equation}
where $l$ and $l_0$ are the energy diffusion lengths in the modulation and the heating channel, respectively.
As explained in Sec.~\ref{bound}, not only $T$ but also $\kappa (\partial T/\partial y)$ has to be continuous.
Making use of Eqs.~(\ref{kappa}) and (\ref{ediff}), we obtain
\begin{equation}
	(T_\text{bulk} -T_\text{i}) \epsilon_0^{(1+\alpha_\text{e}-\alpha_\text{i})/2}
	=
	(T_\text{i} -T_0) (\epsilon_0-V_\text{mc})^{(1+\alpha_\text{e}-\alpha_\text{i})/2}
\end{equation}
which is equivalent to Eq.~(\ref{Tinterface}).

Finally, we derive Eq.~(\ref{s_depleted}) for an almost depleted modulation channel, $\epsilon_0-V_\text{mc} \ll \epsilon_0$.
After the replacement $T_\text{i} \approx T_\text{bulk}$, we calculate $s_0$ by making use of Eqs.~(\ref{A1}) and (\ref{A2}) but, in order to take the gate voltage in the modulation channel into account, we replace in Eq.~(\ref{A2}) $\epsilon_0$ by $\epsilon_0-V_\text{mc}$.
This immediately yields Eq.~(\ref{s_depleted}).

\bibliographystyle{apsrev4-1}

\end{document}